\documentclass[conference,10pt,twoside,twocolumn]{IEEEtran}
\IEEEoverridecommandlockouts
\usepackage{cite}
\usepackage[T1]{fontenc}
\usepackage{graphicx}
\usepackage{amssymb}
\usepackage{amsmath}
\usepackage{amsthm}
\usepackage{subfigure}
\usepackage{microtype}
\usepackage{balance}
\usepackage{xcolor}
\usepackage{algorithm}
\usepackage{algorithmicx}
\usepackage{algpseudocode}
\algrenewcommand\algorithmicindent{0.7em}%

\usepackage{amssymb}
\usepackage{amsfonts}
\usepackage{mathrsfs}
\usepackage{xspace}
\usepackage{bm}
\usepackage{upgreek}

\newcommand{\safemath}[2]{\newcommand{#1}{\ensuremath{#2}\xspace}}

\safemath{\bma}{\mathbf{a}}
\safemath{\bmb}{\mathbf{b}}
\safemath{\bmc}{\mathbf{c}}
\safemath{\bmd}{\mathbf{d}}
\safemath{\bme}{\mathbf{e}}
\safemath{\bmf}{\mathbf{f}}
\safemath{\bmg}{\mathbf{g}}
\safemath{\bmh}{\mathbf{h}}
\safemath{\bmi}{\mathbf{i}}
\safemath{\bmj}{\mathbf{j}}
\safemath{\bmk}{\mathbf{k}}
\safemath{\bml}{\mathbf{l}}
\safemath{\bmm}{\mathbf{m}}
\safemath{\bmn}{\mathbf{n}}
\safemath{\bmo}{\mathbf{o}}
\safemath{\bmp}{\mathbf{p}}
\safemath{\bmq}{\mathbf{q}}
\safemath{\bmr}{\mathbf{r}}
\safemath{\bms}{\mathbf{s}}
\safemath{\bmt}{\mathbf{t}}
\safemath{\bmu}{\mathbf{u}}
\safemath{\bmv}{\mathbf{v}}
\safemath{\bmw}{\mathbf{w}}
\safemath{\bmx}{\mathbf{x}}
\safemath{\bmy}{\mathbf{y}}
\safemath{\bmz}{\mathbf{z}}
\safemath{\bmzero}{\mathbf{0}}
\safemath{\bmone}{\mathbf{1}}
\safemath{\Bell}{\ensuremath{\boldsymbol\ell}}

\bmdefine{\biad}{a}
\bmdefine{\bibd}{b}
\bmdefine{\bicd}{c}
\bmdefine{\bidd}{d}
\bmdefine{\bied}{e}
\bmdefine{\bifd}{f}
\bmdefine{\bigd}{g}
\bmdefine{\bihd}{h}
\bmdefine{\biid}{i}
\bmdefine{\bijd}{j}
\bmdefine{\bikd}{k}
\bmdefine{\bild}{l}
\bmdefine{\bimd}{m}
\bmdefine{\bind}{n}
\bmdefine{\biod}{o}
\bmdefine{\bipd}{p}
\bmdefine{\biqd}{q}
\bmdefine{\bird}{r}
\bmdefine{\bisd}{s}
\bmdefine{\bitd}{t}
\bmdefine{\biud}{u}
\bmdefine{\bivd}{v}
\bmdefine{\biwd}{w}
\bmdefine{\bixd}{x}
\bmdefine{\biyd}{y}
\bmdefine{\bizd}{z}

\bmdefine{\bixid}{\xi}
\bmdefine{\bilambdad}{\lambda}
\bmdefine{\bimud}{\mu}
\bmdefine{\bithetad}{\theta}
\bmdefine{\biphid}{\phi}
\bmdefine{\bideltad}{\delta}

\safemath{\bmia}{\biad}
\safemath{\bmib}{\bibd}
\safemath{\bmic}{\bicd}
\safemath{\bmid}{\bidd}
\safemath{\bmie}{\bied}
\safemath{\bmif}{\bifd}
\safemath{\bmig}{\bigd}
\safemath{\bmih}{\bihd}
\safemath{\bmii}{\biid}
\safemath{\bmij}{\bijd}
\safemath{\bmik}{\bikd}
\safemath{\bmil}{\bild}
\safemath{\bmim}{\bimd}
\safemath{\bmin}{\bind}
\safemath{\bmio}{\biod}
\safemath{\bmip}{\bipd}
\safemath{\bmiq}{\biqd}
\safemath{\bmir}{\bird}
\safemath{\bmis}{\bisd}
\safemath{\bmit}{\bitd}
\safemath{\bmiu}{\biud}
\safemath{\bmiv}{\bivd}
\safemath{\bmiw}{\biwd}
\safemath{\bmix}{\bixd}
\safemath{\bmiy}{\biyd}
\safemath{\bmiz}{\bizd}

\safemath{\bmxi}{\bixid}
\safemath{\bmlambda}{\bilambdad}
\safemath{\bmmu}{\bimud}
\safemath{\bmtheta}{\bithetad}
\safemath{\bmphi}{\biphid}
\safemath{\bmdelta}{\bideltad}

\safemath{\bA}{\mathbf{A}}
\safemath{\bB}{\mathbf{B}}
\safemath{\bC}{\mathbf{C}}
\safemath{\bD}{\mathbf{D}}
\safemath{\bE}{\mathbf{E}}
\safemath{\bF}{\mathbf{F}}
\safemath{\bG}{\mathbf{G}}
\safemath{\bH}{\mathbf{H}}
\safemath{\bI}{\mathbf{I}}
\safemath{\bJ}{\mathbf{J}}
\safemath{\bK}{\mathbf{K}}
\safemath{\bL}{\mathbf{L}}
\safemath{\bM}{\mathbf{M}}
\safemath{\bN}{\mathbf{N}}
\safemath{\bO}{\mathbf{O}}
\safemath{\bP}{\mathbf{P}}
\safemath{\bQ}{\mathbf{Q}}
\safemath{\bR}{\mathbf{R}}
\safemath{\bS}{\mathbf{S}}
\safemath{\bT}{\mathbf{T}}
\safemath{\bU}{\mathbf{U}}
\safemath{\bV}{\mathbf{V}}
\safemath{\bW}{\mathbf{W}}
\safemath{\bX}{\mathbf{X}}
\safemath{\bY}{\mathbf{Y}}
\safemath{\bZ}{\mathbf{Z}}

\safemath{\bZero}{\mathbf{0}}
\safemath{\bOne}{\mathbf{1}}
\safemath{\bDelta}{\mathbf{\Delta}}
\safemath{\bLambda}{\mathbf{\UpLambda}}
\safemath{\bPhi}{\mathbf{\Upphi}}
\safemath{\bSigma}{\mathbf{\Upsigma}}
\safemath{\bOmega}{\mathbf{\Upomega}}
\safemath{\bTheta}{\mathbf{\Uptheta}}

\bmdefine{\biAd}{A}
\bmdefine{\biBd}{B}
\bmdefine{\biCd}{C}
\bmdefine{\biDd}{D}
\bmdefine{\biEd}{E}
\bmdefine{\biFd}{F}
\bmdefine{\biGd}{G}
\bmdefine{\biHd}{H}
\bmdefine{\biId}{I}
\bmdefine{\biJd}{J}
\bmdefine{\biKd}{K}
\bmdefine{\biLd}{L}
\bmdefine{\biMd}{M}
\bmdefine{\biOd}{N}
\bmdefine{\biPd}{O}
\bmdefine{\biQd}{P}
\bmdefine{\biRd}{R}
\bmdefine{\biSd}{S}
\bmdefine{\biTd}{T}
\bmdefine{\biUd}{U}
\bmdefine{\biVd}{V}
\bmdefine{\biWd}{W}
\bmdefine{\biXd}{X}
\bmdefine{\biYd}{Y}
\bmdefine{\biZd}{Z}

\bmdefine{\biDelta}{\Delta}
\bmdefine{\biLambda}{\Lambda}
\bmdefine{\biPhi}{\Phi}
\bmdefine{\biSigma}{\Sigma}
\bmdefine{\biOmega}{\Omega}
\bmdefine{\biTheta}{\Theta}

\safemath{\bimA}{\biAd}
\safemath{\bimB}{\biBd}
\safemath{\bimC}{\biCd}
\safemath{\bimD}{\biDd}
\safemath{\bimE}{\biEd}
\safemath{\bimF}{\biFd}
\safemath{\bimG}{\biGd}
\safemath{\bimH}{\biHd}
\safemath{\bimI}{\biId}
\safemath{\bimJ}{\biJd}
\safemath{\bimK}{\biKd}
\safemath{\bimL}{\biLd}
\safemath{\bimM}{\biMd}
\safemath{\bimN}{\biNd}
\safemath{\bimO}{\biOd}
\safemath{\bimP}{\biPd}
\safemath{\bimQ}{\biQd}
\safemath{\bimR}{\biRd}
\safemath{\bimS}{\biSd}
\safemath{\bimT}{\biTd}
\safemath{\bimU}{\biUd}
\safemath{\bimV}{\biVd}
\safemath{\bimW}{\biWd}
\safemath{\bimX}{\biXd}
\safemath{\bimY}{\biYd}
\safemath{\bimZ}{\biZd}

\safemath{\bimDelta}{\biDelta}
\safemath{\bimLambda}{\biLambda}
\safemath{\bimPhi}{\biPhi}
\safemath{\bimSigma}{\biSigma}
\safemath{\bimOmega}{\biOmega}
\safemath{\bimTheta}{\biTheta}

\safemath{\setA}{\mathcal{A}}
\safemath{\setB}{\mathcal{B}}
\safemath{\setC}{\mathcal{C}}
\safemath{\setD}{\mathcal{D}}
\safemath{\setE}{\mathcal{E}}
\safemath{\setF}{\mathcal{F}}
\safemath{\setG}{\mathcal{G}}
\safemath{\setH}{\mathcal{H}}
\safemath{\setI}{\mathcal{I}}
\safemath{\setJ}{\mathcal{J}}
\safemath{\setK}{\mathcal{K}}
\safemath{\setL}{\mathcal{L}}
\safemath{\setM}{\mathcal{M}}
\safemath{\setN}{\mathcal{N}}
\safemath{\setO}{\mathcal{O}}
\safemath{\setP}{\mathcal{P}}
\safemath{\setQ}{\mathcal{Q}}
\safemath{\setR}{\mathcal{R}}
\safemath{\setS}{\mathcal{S}}
\safemath{\setT}{\mathcal{T}}
\safemath{\setU}{\mathcal{U}}
\safemath{\setV}{\mathcal{V}}
\safemath{\setW}{\mathcal{W}}
\safemath{\setX}{\mathcal{X}}
\safemath{\setY}{\mathcal{Y}}
\safemath{\setZ}{\mathcal{Z}}
\safemath{\emptySet}{\varnothing}

\safemath{\colA}{\mathscr{A}}
\safemath{\colB}{\mathscr{B}}
\safemath{\colC}{\mathscr{C}}
\safemath{\colD}{\mathscr{D}}
\safemath{\colE}{\mathscr{E}}
\safemath{\colF}{\mathscr{F}}
\safemath{\colG}{\mathscr{G}}
\safemath{\colH}{\mathscr{H}}
\safemath{\colI}{\mathscr{I}}
\safemath{\colJ}{\mathscr{J}}
\safemath{\colK}{\mathscr{K}}
\safemath{\colL}{\mathscr{L}}
\safemath{\colM}{\mathscr{M}}
\safemath{\colN}{\mathscr{N}}
\safemath{\colO}{\mathscr{O}}
\safemath{\colP}{\mathscr{P}}
\safemath{\colQ}{\mathscr{Q}}
\safemath{\colR}{\mathscr{R}}
\safemath{\colS}{\mathscr{S}}
\safemath{\colT}{\mathscr{T}}
\safemath{\colU}{\mathscr{U}}
\safemath{\colV}{\mathscr{V}}
\safemath{\colW}{\mathscr{W}}
\safemath{\colX}{\mathscr{X}}
\safemath{\colY}{\mathscr{Y}}
\safemath{\colZ}{\mathscr{Z}}

\safemath{\opA}{\mathbb{A}}
\safemath{\opB}{\mathbb{B}}
\safemath{\opC}{\mathbb{C}}
\safemath{\opD}{\mathbb{D}}
\safemath{\opE}{\mathbb{E}}
\safemath{\opF}{\mathbb{F}}
\safemath{\opG}{\mathbb{G}}
\safemath{\opH}{\mathbb{H}}
\safemath{\opI}{\mathbb{I}}
\safemath{\opJ}{\mathbb{J}}
\safemath{\opK}{\mathbb{K}}
\safemath{\opL}{\mathbb{L}}
\safemath{\opM}{\mathbb{M}}
\safemath{\opN}{\mathbb{N}}
\safemath{\opO}{\mathbb{O}}
\safemath{\opP}{\mathbb{P}}
\safemath{\opQ}{\mathbb{Q}}
\safemath{\opR}{\mathbb{R}}
\safemath{\opS}{\mathbb{S}}
\safemath{\opT}{\mathbb{T}}
\safemath{\opU}{\mathbb{U}}
\safemath{\opV}{\mathbb{V}}
\safemath{\opW}{\mathbb{W}}
\safemath{\opX}{\mathbb{X}}
\safemath{\opY}{\mathbb{Y}}
\safemath{\opZ}{\mathbb{Z}}
\safemath{\opZero}{\mathbb{O}}
\safemath{\identityop}{\opI}

\safemath{\veca}{\bma}
\safemath{\vecb}{\bmb}
\safemath{\vecc}{\bmc}
\safemath{\vecd}{\bmd}
\safemath{\vece}{\bme}
\safemath{\vecf}{\bmf}
\safemath{\vecg}{\bmg}
\safemath{\vech}{\bmh}
\safemath{\veci}{\bmi}
\safemath{\vecj}{\bmj}
\safemath{\veck}{\bmk}
\safemath{\vecl}{\bml}
\safemath{\vecm}{\bmm}
\safemath{\vecn}{\bmn}
\safemath{\veco}{\bmo}
\safemath{\vecp}{\bmp}
\safemath{\vecq}{\bmq}
\safemath{\vecr}{\bmr}
\safemath{\vecs}{\bms}
\safemath{\vect}{\bmt}
\safemath{\vecu}{\bmu}
\safemath{\vecv}{\bmv}
\safemath{\vecw}{\bmw}
\safemath{\vecx}{\bmx}
\safemath{\vecy}{\bmy}
\safemath{\vecz}{\bmz}

\safemath{\veczero}{\bmzero}
\safemath{\vecone}{\bmone}
\safemath{\vecxi}{\bmxi}
\safemath{\veclambda}{\bmlambda}
\safemath{\vecmu}{\bmmu}
\safemath{\vectheta}{\bmtheta}
\safemath{\vecphi}{\bmphi}
\safemath{\vecdelta}{\bmdelta}

\safemath{\matA}{\bA}
\safemath{\matB}{\bB}
\safemath{\matC}{\bC}
\safemath{\matD}{\bD}
\safemath{\matE}{\bE}
\safemath{\matF}{\bF}
\safemath{\matG}{\bG}
\safemath{\matH}{\bH}
\safemath{\matI}{\bI}
\safemath{\matJ}{\bJ}
\safemath{\matK}{\bK}
\safemath{\matL}{\bL}
\safemath{\matM}{\bM}
\safemath{\matN}{\bN}
\safemath{\matO}{\bO}
\safemath{\matP}{\bP}
\safemath{\matQ}{\bQ}
\safemath{\matR}{\bR}
\safemath{\matS}{\bS}
\safemath{\matT}{\bT}
\safemath{\matU}{\bU}
\safemath{\matV}{\bV}
\safemath{\matW}{\bW}
\safemath{\matX}{\bX}
\safemath{\matY}{\bY}
\safemath{\matZ}{\bZ}
\safemath{\matzero}{\bmzero}

\safemath{\matDelta}{\bDelta}
\safemath{\matLambda}{\bLambda}
\safemath{\matPhi}{\bPhi}
\safemath{\matSigma}{\bSigma}
\safemath{\matOmega}{\bOmega}
\safemath{\matTheta}{\bTheta}

\safemath{\matidentity}{\matI}
\safemath{\matone}{\matO}

\safemath{\rnda}{A}
\safemath{\rndb}{B}
\safemath{\rndc}{C}
\safemath{\rndd}{D}
\safemath{\rnde}{E}
\safemath{\rndf}{F}
\safemath{\rndg}{G}
\safemath{\rndh}{H}
\safemath{\rndi}{I}
\safemath{\rndj}{J}
\safemath{\rndk}{K}
\safemath{\rndl}{L}
\safemath{\rndm}{M}
\safemath{\rndn}{N}
\safemath{\rndo}{O}
\safemath{\rndp}{P}
\safemath{\rndq}{Q}
\safemath{\rndr}{R}
\safemath{\rnds}{S}
\safemath{\rndt}{T}
\safemath{\rndu}{U}
\safemath{\rndv}{V}
\safemath{\rndw}{W}
\safemath{\rndx}{X}
\safemath{\rndy}{Y}
\safemath{\rndz}{Z}

\safemath{\rveca}{\bimA}
\safemath{\rvecb}{\bimB}
\safemath{\rvecc}{\bimC}
\safemath{\rvecd}{\bimD}
\safemath{\rvece}{\bimE}
\safemath{\rvecf}{\bimF}
\safemath{\rvecg}{\bimG}
\safemath{\rvech}{\bimH}
\safemath{\rveci}{\bimI}
\safemath{\rvecj}{\bimJ}
\safemath{\rveck}{\bimK}
\safemath{\rvecl}{\bimL}
\safemath{\rvecm}{\bimM}
\safemath{\rvecn}{\bimN}
\safemath{\rveco}{\bomO}
\safemath{\rvecp}{\bimP}
\safemath{\rvecq}{\bimQ}
\safemath{\rvecr}{\bimR}
\safemath{\rvecs}{\bimS}
\safemath{\rvect}{\bimT}
\safemath{\rvecu}{\bimU}
\safemath{\rvecv}{\bimV}
\safemath{\rvecw}{\bimW}
\safemath{\rvecx}{\bimX}
\safemath{\rvecy}{\bimY}
\safemath{\rvecz}{\bimZ}

\safemath{\rvecxi}{\bmxi}
\safemath{\rveclambda}{\bmlambda}
\safemath{\rvecmu}{\bmmu}
\safemath{\rvectheta}{\bmtheta}
\safemath{\rvecphi}{\bmphi}

\safemath{\rmatA}{\bimA}
\safemath{\rmatB}{\bimB}
\safemath{\rmatC}{\bimC}
\safemath{\rmatD}{\bimD}
\safemath{\rmatE}{\bimE}
\safemath{\rmatF}{\bimF}
\safemath{\rmatG}{\bimG}
\safemath{\rmatH}{\bimH}
\safemath{\rmatI}{\bimI}
\safemath{\rmatJ}{\bimJ}
\safemath{\rmatK}{\bimK}
\safemath{\rmatL}{\bimL}
\safemath{\rmatM}{\bimM}
\safemath{\rmatN}{\bimN}
\safemath{\rmatO}{\bimO}
\safemath{\rmatP}{\bimP}
\safemath{\rmatQ}{\bimQ}
\safemath{\rmatR}{\bimR}
\safemath{\rmatS}{\bimS}
\safemath{\rmatT}{\bimT}
\safemath{\rmatU}{\bimU}
\safemath{\rmatV}{\bimV}
\safemath{\rmatW}{\bimW}
\safemath{\rmatX}{\bimX}
\safemath{\rmatY}{\bimY}
\safemath{\rmatZ}{\bimZ}

\safemath{\rmatDelta}{\bimDelta}
\safemath{\rmatLambda}{\bimLambda}
\safemath{\rmatPhi}{\bimPhi}
\safemath{\rmatSigma}{\bimSigma}
\safemath{\rmatOmega}{\bimOmega}
\safemath{\rmatTheta}{\bimTheta}

\usepackage{amssymb}
\usepackage{amsfonts}
\usepackage{mathrsfs}
\usepackage{xspace}
\usepackage{bm}
\usepackage{fancyref}
\usepackage{textcomp}

\usepackage{multirow}
\usepackage{stmaryrd}

\newenvironment{textbmatrix}{	\setlength{\arraycolsep}{2.5pt}%
								\left[\begin{matrix}}{\end{matrix}\right]%
								\raisebox{0.08ex}{\vphantom{M}}}

\def\be{\begin{equation}}
\def\ee{\end{equation}}
\def\een{\nonumber \end{equation}}
\def\mat{\begin{bmatrix}}
\def\emat{\end{bmatrix}}
\def\btm{\begin{textbmatrix}}
\def\etm{\end{textbmatrix}}

\def\ba#1\ea{\begin{align}#1\end{align}}
\def\bas#1\eas{\begin{align*}#1\end{align*}}
\def\bs#1\es{\begin{split}#1\end{split}}
\def\bg#1\eg{\begin{gather}#1\end{gather}}
\def\bml#1\eml{\begin{multline}#1\end{multline}}
\def\bi#1\ei{\begin{itemize}#1\end{itemize}}

\DeclareMathOperator{\Prob}{\opP}			

\safemath{\dirac}{\delta}					
\safemath{\krond}{\dirac}					

\safemath{\upto}{\uparrow}
\safemath{\downto}{\downarrow}
\safemath{\iu}{j}							
\safemath{\ev}{\lambda}						
\safemath{\hilseqspace}{l^{2}}				
\newcommand{\banachfunspace}[1]{\setL^{#1}}	
\safemath{\hilfunspace}{\banachfunspace{2}}	

\safemath{\SNR}{\textit{SNR}} 				
\safemath{\PAR}{\textit{PAR}} 				
\safemath{\No}{N_0}							
\safemath{\Es}{E_s}							
\safemath{\Eb}{E_b}							
\safemath{\EbNo}{\frac{\Eb}{\No}}
\safemath{\EsNo}{\frac{\Es}{\No}}

\DeclareMathOperator{\CHop}{\ensuremath{\opH}} 
\safemath{\tvir}{\rndh_{\CHop}}				
\safemath{\tvtf}{\rndl_{\CHop}}				
\safemath{\spf}{\rnds_{\CHop}}				
\safemath{\bff}{H_{\CHop}}					

\safemath{\ircf}{r_{h}}						
\safemath{\tftvcf}{r_{s}}					
\safemath{\tfcf}{r_{l}}						
\safemath{\bfcf}{r_{H}}						

\safemath{\tcorr}{c_h}						
\safemath{\scf}{c_{s}}						
\safemath{\tfcorr}{c_{l}}					
\safemath{\fcorr}{c_{H}}						

\safemath{\mi}{I}							
\safemath{\capacity}{C}						

\safemath{\normal}{\mathcal{N}}			
\safemath{\jpg}{\mathcal{CN}}			
\safemath{\mchain}{\leftrightarrow}		

\safemath{\dB}{\,\mathrm{dB}}
\safemath{\dBm}{\,\mathrm{dBm}}
\safemath{\Hz}{\,\mathrm{Hz}}
\safemath{\kHz}{\,\mathrm{kHz}}
\safemath{\MHz}{\,\mathrm{MHz}}
\safemath{\GHz}{\,\mathrm{GHz}}
\safemath{\s}{\,\mathrm{s}}
\safemath{\ms}{\,\mathrm{ms}}
\safemath{\mus}{\,\mathrm{\text{\textmu}s}}
\safemath{\ns}{\,\mathrm{ns}}
\safemath{\ps}{\,\mathrm{ps}}
\safemath{\meter}{\,\mathrm{m}}
\safemath{\mm}{\,\mathrm{mm}}
\safemath{\cm}{\,\mathrm{cm}}
\safemath{\m}{\,\mathrm{m}}
\safemath{\W}{\,\mathrm{W}}
\safemath{\mW}{\, \mathrm{mW}}
\safemath{\J}{\,\mathrm{J}}
\safemath{\K}{\,\mathrm{K}}
\safemath{\bit}{\,\mathrm{bit}}
\safemath{\nat}{\,\mathrm{nat}}

\safemath{\define}{\triangleq}			

\safemath{\equivalent}{\sim}
\safemath{\distas}{\sim}					
\safemath{\sdiff}{\Delta}				

\safemath{\reals}{\mathbb{R}}
\safemath{\positivereals}{\reals_{+}}
\safemath{\integers}{\mathbb{Z}}
\safemath{\posint}{\integers_{+}}
\safemath{\naturals}{\mathbb{N}}
\safemath{\posnaturals}{\naturals_{+}}
\safemath{\complexset}{\mathbb{C}}
\safemath{\rationals}{\mathbb{Q}}

\newcommand*{\fancyrefapplabelprefix}{app}		
\newcommand*{\fancyrefthmlabelprefix}{thm}		
\newcommand*{\fancyreflemlabelprefix}{lem}		
\newcommand*{\fancyrefcorlabelprefix}{cor}		
\newcommand*{\fancyrefdeflabelprefix}{def}		
\newcommand*{\fancyrefproplabelprefix}{prop}		
\newcommand*{\fancyrefexmpllabelprefix}{exmpl}
\newcommand*{\fancyrefalglabelprefix}{alg}		
\newcommand*{\fancyreftbllabelprefix}{tbl}		

\frefformat{vario}{\fancyrefseclabelprefix}{Sec.~#1}
\frefformat{vario}{\fancyrefthmlabelprefix}{Thm.~#1}
\frefformat{vario}{\fancyreftbllabelprefix}{Tbl.~#1}
\frefformat{vario}{\fancyreflemlabelprefix}{Lem.~#1}
\frefformat{vario}{\fancyrefcorlabelprefix}{Corr.~#1}
\frefformat{vario}{\fancyrefdeflabelprefix}{Def.~#1}
\frefformat{vario}{\fancyreffiglabelprefix}{Fig.~#1}
\frefformat{vario}{\fancyrefapplabelprefix}{App.~#1}
\frefformat{vario}{\fancyrefeqlabelprefix}{(#1)}
\frefformat{vario}{\fancyrefproplabelprefix}{Prop.~#1}
\frefformat{vario}{\fancyrefexmpllabelprefix}{Ex.~#1}
\frefformat{vario}{\fancyrefalglabelprefix}{Alg.~#1}

\safemath{\dictab}{[\,\dicta\,\,\dictb\,]}

\safemath{\ysig}{\bmy}
\safemath{\ysighat}{\hat{\ysig}}
\safemath{\ysigdim}{M}
\safemath{\xsig}{\bmx}
\safemath{\xsigdim}{N}
\safemath{\nx}{n_x}
\safemath{\zsig}{\bmz}
\safemath{\zsigdim}{\ysigdim}
\safemath{\rsig}{\bmr}
\safemath{\Adict}{\bA}
\safemath{\Adicttilde}{\widetilde{\Adict}}
\safemath{\Adictdim}{\outputdim\times\xsigdim}
\safemath{\avec}{\bma}
\safemath{\avectilde}{\tilde{\avec}}
\safemath{\Bdict}{\bB}
\safemath{\Bdicttilde}{\widetilde{\Bdict}}
\safemath{\Cdict}{\bC}
\safemath{\cvec}{\bmc}
\safemath{\Ddict}{\bD}
\safemath{\Ddictdim}{\ysigdim\times\xsigdim}
\safemath{\dvec}{\bmd}
\safemath{\Ddicttilde}{\widetilde{\bD}}
\safemath{\Bonb}{\bB}
\safemath{\bvec}{\bmb}
\safemath{\Bonbdim}{\ysigdim\times\ysigdim}
\safemath{\noise}{\bmn}
\safemath{\noisedim}{\ysigim}
\safemath{\err}{\bme}
\safemath{\errdim}{\ysigdim}
\safemath{\errset}{\setE}
\safemath{\nerr}{n_e}
\safemath{\delop}{\bP_\errset}
\safemath{\delopc}{\bP_{{\errset}^c}}

\safemath{\cplxi}{\imath}
\safemath{\cplxj}{\jmath}

\safemath{\dict}{\matD}
\safemath{\inputdim}{N}		
\safemath{\outputdim}{M}		
\safemath{\sparsity}{S}	
\safemath{\inputdimA}{{N_a}}	
\safemath{\inputdimB}{{N_b}}	
\safemath{\elemA}{{n_a}}	
\safemath{\elemB}{{n_b}}	
\safemath{\resA}{\matR_a}	
\safemath{\resB}{\matR_b}	
\safemath{\subD}{\matS} 
\safemath{\subA}{\matS_a} 
\safemath{\subB}{\matS_b} 
\safemath{\dicta}{\matA} 	
\safemath{\dictb}{\matB} 	
\safemath{\hollowS}{H}
\safemath{\hollowA}{H_a}
\safemath{\hollowB}{H_b}
\safemath{\cross}{Z}
\safemath{\coh}{\mu_d}			
\safemath{\coha}{\mu_a}			
\safemath{\cohb}{\mu_b}			
\safemath{\mubs}{\nu}	
\safemath{\cohm}{\mu_m} 
\safemath{\dictset}{\setD}	
\safemath{\dictsetp}{\dictset(\coh,\coha,\cohb)}	
\safemath{\dictsetgen}{\dictset_\text{gen}}
\safemath{\dictsetgenp}{\dictsetgen(\coh)}
\safemath{\dictsetonb}{\dictset_\text{onb}}
\safemath{\dictsetonbp}{\dictsetonb(\coh)}

\safemath{\leftside}{U}
\safemath{\rightsideA}{R_a}
\safemath{\rightsideB}{R_b}

\safemath{\indexS}{\setI_S} 

\safemath{\na}{n_a}			
\safemath{\nb}{n_b}			
\safemath{\coeffa}{p_i}	
\safemath{\coeffb}{q_j}	
\safemath{\seta}{\setP}		
\safemath{\setb}{\setQ}     
\safemath{\setw}{\setW}	
\safemath{\setz}{\setZ}	
\safemath{\cola}{\veca}		
\safemath{\colb}{\vecb}		
\safemath{\cold}{\vecd}		
\safemath{\inputvec}{\vecx} 	
\safemath{\error}{\vece}	
\safemath{\noiseout}{\vecz} 	
\safemath{\inputvecel}{x}
\safemath{\inputveca}{\vecx_a}
\safemath{\inputvecb}{\vecx_b}
\safemath{\outputvec}{\vecy}	
\safemath{\lambdamin}{\lambda_{\mathrm{min}}}

\safemath{\elltwo}{\ell_2}
\safemath{\ellone}{\ell_1}
\safemath{\ellzero}{\ell_0}
\safemath{\ellinf}{\ell_\infty}
\safemath{\ellinftilde}{\ell_{\widetilde\infty}}
\safemath{\licard}{Z(\coh,\coha,\cohb)}
\safemath{\xsol}{\hat{x}}
\safemath{\xbord}{x_b}		
\safemath{\xstat}{x_s}		
\safemath{\xstatLone}{\tilde{x}_s}
\safemath{\order}{\mathcal{O}} 
\safemath{\scales}{\Theta} 
\safemath{\ones}{\mathbf{1}} 
\safemath{\zeroes}{\mathbf{0}} 
\safemath{\thlone}{\kappa(\coh,\cohb)} 
\safemath{\constoneA}{\delta} 
\safemath{\constoneB}{\epsilon} 
\safemath{\nlarge}{L}				   
\safemath{\sumlarge}{S_\nlarge}
\safemath{\maxlarger}{P_\nlarge}	   
\safemath{\Pzero}{\textrm{P0}}	
\safemath{\Pone}{\textrm{P1}}
\safemath{\vecfir}{\vecw}			 
\safemath{\vecsec}{\vecz}
\safemath{\elvecfir}{w}              
\safemath{\elvecsec}{z}				 
\safemath{\nlargefir}{n}
\safemath{\normout}{\gamma}
\safemath{\auxfun}{h}
\safemath{\supp}{\textrm{supp}}

\safemath{\indexa}{\ell}
\safemath{\indexb}{r}
\safemath{\indexc}{i}
\safemath{\indexd}{j}

\safemath{\project}{P}

\safemath{\firstslotset}{\setU_1}  
\safemath{\secondslotset}{\setU_2} 
\safemath{\randomset}{\setS} 

\usepackage{framed}

\safemath{\Tran}{\textnormal{T}}
\safemath{\Herm}{\textnormal{H}}

\newcommand*{\fancyreflstlabelprefix}{lst}
\fancyrefaddcaptions{english}{%
  \providecommand*{\freflstname}{Listing}%
}
\frefformat{vario}{\fancyreflstlabelprefix}{%
  \freflstname\fancyrefdefaultspacing#1#3%
}

\begin{document}
\title{Optimizing  Puncturing Patterns of 5G NR \\ LDPC Codes for Few-Iteration Decoding}
\author{%
    \IEEEauthorblockN{Reinhard Wiesmayr$^\text{1}$, Darja Nonaca$^\text{1}$, Chris Dick$^\text{2}$, and Christoph Studer$^\text{1}$}\\
    \textit{$^\textnormal{1}$ETH Zurich, $^\textnormal{2}$NVIDIA}\\
    \textit{email: \{wiesmayr, dnonaca\}@iis.ee.ethz.ch, cdick@nvidia.com, studer@ethz.ch}
    \thanks{The work of CS was supported in part by an ETH Zurich Research Grant. We acknowledge NVIDIA for their sponsorship of this research.}
    \thanks{The authors thank Nihat Engin Tunali for helpful comments and suggestions.}
    }

\maketitle

\begin{abstract}
Rate-matching of low-density parity-check (LDPC) codes enables a single code description to support a wide range of code lengths and rates. In 5G NR, rate matching is accomplished by extending (lifting) a base code to a desired target length and by puncturing (not transmitting) certain code bits.
LDPC codes and rate matching are typically designed for the asymptotic performance limit with an ideal decoder. 
Practical LDPC decoders, however, carry out tens or fewer message-passing decoding iterations to achieve the target throughput and latency of modern wireless systems. 
We show that one can optimize LDPC code puncturing patterns for such few-iteration-constrained decoders using a method we call swapping of punctured and transmitted blocks (SPAT). 
Our simulation results show that SPAT yields from 0.20\,dB up to 0.55\,dB improved signal-to-noise ratio performance compared to the standard 5G NR LDPC code puncturing pattern for a wide range of code lengths and rates.

\end{abstract}

\section{Introduction}

Digital communication systems rely on channel coding to protect the transmitted data against transmission errors.
Modern channel codes, such as the 5G new radio (NR) low-density parity-check (LDPC) code \cite{ETSI5G}, support a wide range of code rates and lengths using a single, unified code description. 
A common technique for constructing such flexible LDPC codes is \emph{puncturing}, which selects a predetermined set of code bits that are not transmitted.
While punctured LDPC codes have been shown to approach the channel capacity~\cite{hsu2008capacity}, they are usually designed for the asymptotic performance limit~\cite{richardson2018design}, which describes the error-rate performance that could be achieved with an infinitely long code and an ideal decoder that performs infinitely many message-passing (MP) iterations.

MP is the de-facto standard decoding method for practical LDPC codes~\cite{kschischang2001factor} and iteratively propagates reliability information on the coded bits over a bipartite graph described by the code.
Although LDPC codes and puncturing patterns are generally designed for the asymptotic performance limit, practical LDPC codes have finite code-length (e.g., a few thousand code bits or less), and LDPC decoders typically only carry out a small number of MP iterations (e.g., tens or even fewer) in order to meet the stringent throughput, latency, silicon area, and power constraints of modern wireless systems.
Thus, practical decoder implementations operate far from that of an ideal decoder in the asymptotic performance limit.

\subsection{Contributions}

We demonstrate that, with practical MP-based decoders, better puncturing patterns for the 5G-NR-complaint LDPC code exist. 
To this end, we propose SPAT (short for Swapping of Punctured And Transmitted blocks), an algorithm that optimizes LDPC code puncturing patterns for few-iteration MP decoding.
SPAT delivers puncturing patterns that achieve a block error rate (BLER) of $10^{-3}$ at up to 0.55\,dB lower signal-to-noise ratio (SNR) than the standard 5G NR LDPC code for varying code lengths and rates as well as for flooding and layered MP decoding. 
These gains are achieved solely by optimizing the puncturing patterns while leaving the remaining 5G NR rate-matching scheme and the underlying 5G NR LDPC base code unchanged.
To gain insight into the inner workings of such optimized puncturing patterns, we conduct a performance analysis based on the empirical average binary cross-entropy (BCE). This analysis reveals that optimized puncturing patterns achieve lower error rates for fewer MP iterations, but can be outperformed by the 5G-NR-compliant puncturing pattern for MP-based decoders that carry out a large number of iterations. 

\subsection{Relevant Prior Work}

LDPC codes, discovered by Gallager in \cite{gallager1962low}, are widely used in modern communication systems due to their excellent error-correction capabilities and efficient decoder implementations. 
State-of-the-art LDPC codes~\cite{richardson2018design}, including the code specified in 5G NR \cite{ETSI5G}, are typically designed with density evolution techniques \cite{richardson2001design} or EXIT (short for extrinsic information transfer) chart methods \cite{ten2004design} that consider 
the asymptotic performance limit with an infinitely long code and an ideal decoder that performs infinitely many iterations.
However, practical LDPC decoder implementations for modern wireless systems must meet stringent throughput, latency, silicon area, and power constraints, which is achieved by carrying out a small number of decoding iterations. 
In particular, real-world MP-based LDPC decoder implementations often utilize $N_{\textnormal{F}}=10$ iterations for the flooding schedule proposed in \cite{gallager1962low} (see, e.g., the implementation in  \cite{milicevic2018multi, park, weiner} )
and $N_{\textnormal{L}}=5$ iterations for the layered schedule proposed in 
\cite{sharon2007efficient} (see, e.g., the implementations in \cite{studer2008configurable,ren2024generalized, lin202133}). 
Evidently, such real-world LDPC decoder implementations operate far away from the asymptotic performance limit. 
In order to achieve improved error-correction performance with such practical few-iteration LDPC decoders, we utilize SPAT to optimize puncturing patterns for $N_{\textnormal{F}}=10$ flooding iterations, which also performs well for $N_{\textnormal{L}}=5$ layered iterations. 

When few-iteration decoding is considered already during code design, better codes can be obtained. This observation was made by Elkelesh \emph{et al.} in \cite{elkelesh2019decoder}, where codes were generated for decoders running $N_{\textnormal{F}}=20$ flooding iterations. 
Their code-construction method was used to generate a short rate-half LDPC code with code-length $128$, which achieves more than $0.3$\,dB better SNR performance at a target block error rate of $10^{-5}$ compared to the rate-matched 5G NR code of the same rate and length in a {real-valued} binary-input additive white Gaussian noise (BI-AWGN) channel.
While the method from~\cite{elkelesh2019decoder} specifically designs a code for one length and rate, an extension to variable code lengths using lifting was presented in \cite{liu2021genalg}.\footnote{State-of-the-art lifting techniques (see, e.g., \cite{myung2006lifting}) do not focus on the error-rate performance with few-iteration decoding and routinely ignore that practical decoders are limited to carry out a small number of MP iterations.}
However, both design methods \cite{elkelesh2019decoder, liu2021genalg} have not yet been applied for generating rate-compatible LDPC codes, i.e., codes that support varying rates. The missing part for a rate adaptive code is puncturing, which truncates the LDPC code to a shorter length by not transmitting certain bits, which increases the rate as the encoded information remains unchanged.

While puncturing patterns (i.e., the selection of bits that are not transmitted) have been optimized for few-iteration decoding already with structured block-type LDPC codes~\cite{hong2006optimal}, existing methods for general LDPC codes optimize the performance in the asymptotic limit and ignore finite (and small) iteration effects.
Prominent puncturing pattern optimization techniques are proposed in, e.g., \cite{ha2004rate} and \cite{LIU20151582}. Reference~\cite{ha2004rate} applies a Gaussian approximation to optimize the punctured code's performance in the asymptotic decoding limit. Reference~\cite{LIU20151582} proposes three optimization methods to puncture short quasi-cyclic (QC)-LDPC codes, but without an explicit focus on few-iteration decoding. The first two methods optimize puncturing based on cycles in the bipartite graph describing the code and MP decoder. The third method randomly generates puncturing patterns and selects the one achieving best error-rate performance using Monte--Carlo simulations. Different from the methods presented in \cite{LIU20151582}, SPAT implements a Greedy algorithm that successively improves an existing puncturing pattern, in our case the 5G-NR-complaint puncturing pattern. Another key difference to our work is our focus on the practical few-iteration decoding performance, which we explicitly study for a varying number of MP iterations.\footnote{The numerical optimization heuristic and also the results in \cite{LIU20151582} are simulated with a decoder that implements 60 or 200 MP iterations, which is far from what practical decoder implementations typically carry out.}
We thereby show that much better few-iteration performance can be achieved just by optimizing the puncturing pattern of an existing code that was designed for the asymptotic performance limit.

\subsection{Notation}

Boldface lowercase and uppercase letters represent column vectors and matrices, respectively. 
For a vector $\bmx$, the $i$th entry is denoted by $x_i = [\bmx]_i$ and its transpose by $\bmx^\Tran$. 
Vertical concatenation of two vectors (or matrices) $\bmx$ and $\bmy$ is denoted by $[\bmx; \bmy]$.
The set $\{0,1\}^n$ indicates all $2^{n}$ vectors of length $n$ with entries taken from the set $\{0,1\}$.
The $M\times N$ all-zeros and $N\times N$ identity matrix are $\bZero_{M\times N}$ and~$\bI_N$, respectively.
GF(2) is the finite field of two elements ($0$ and $1$). 
We denote probabilities by $\Prob(\cdot)$ and conditional probabilities by $\Prob(\cdot|\cdot)$.
\section{Prerequisites}\label{sec:rerequisites}

We now briefly review the basics of the 5G NR LDPC code and the used channel model. Our optimization of puncturing patterns is detailed in \fref{sec:optimization}.

\subsection{Channel Coding and the 5G NR LDPC Code}\label{sec:rate_matching_5g}

In what follows, we focus on the 5G NR LDPC code \cite{ETSI5G}, which is a quasi-cyclic (QC) code in GF(2) that is constructed from a short base code described by a so called protograph, or base graph. 
At the transmitter, $k$ information bits $\vecb\in\{0,1\}^k$ are mapped to a codeword $\bmc\in\setC$ from the finite codebook
\begin{align}
    \setC = \big\{\vecc\in\{0,1\}^n : \bH\bmc = \bZero_{(n-k)\times1}\big\},\label{eq:codebook}
\end{align}
where $\bH$ is the $(n-k)\times n$ parity-check matrix. This mapping is typically implemented by the encoding function $\vecc=\textnormal{enc}(\vecb) = \matG \vecb$, with the $n\times k$ generator matrix $\matG=[\bI_k; \matP]$, that can be obtained from $\bH$. The first $k$ bits in $\vecc$ are the \emph{systematic} bits and the $n-k$ bits generated by $\matP$ are the \emph{parity} bits. 

\begin{figure}[tp]
    \centering
    \includegraphics[width=1.0\columnwidth]{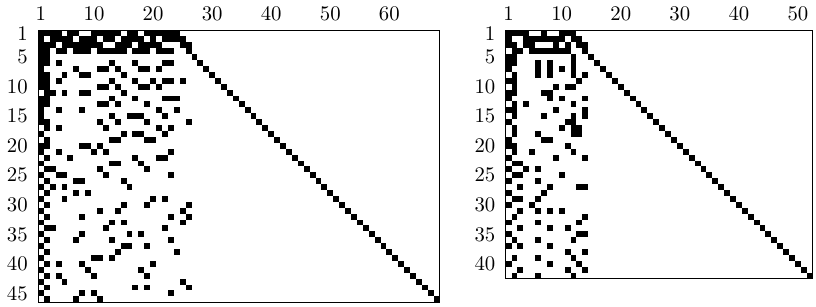}
    \caption{%
    Base graph 1 of dimension $46\times68$ (left) and base graph 2 of dimension $42\times52$ (right). White represents zeros in the BG and will be replaced by zero matrices during lifting. The black boxes represent ones in the BG and will be replaced by circularly shifted identity matrices during lifting. The 5G NR standard specifies a set of shift coefficients for each one-entry in the BG.}
    \label{fig:bg}
\end{figure}

To obtain a parity check matrix for a code of length $n$ and rate $r=k/n$, the 5G NR standard utilizes \emph{rate-matching}, which comprises the following steps: 

First, one selects a base graph (BG), which is a $(n_{\textnormal{B}}-k_{\textnormal{B}})\times n_{\textnormal{B}}$ parity-check matrix that specifies the protograph of the code, and the lifting coefficients, which we explain below. 
In 5G NR, one selects among BG\,1 (of size $46\times68$) and BG\,2 (of size $42\times52$). The structure of the two BGs is illustrated in \fref{fig:bg} and the selection for either BG 1 or BG 2 depends on the desired number of information bits $k$ and the desired rate~$r$. 
Note that the first two columns of both BGs are highly connected to many parity check equations and receive special attention in the design of the 5G NR LDPC code \cite{richardson2018design}.

Second, \emph{lifting} (expanding) the BG is used to increase the length of the base-code (i.e., the code specified by the protograph). 
The selected BG is lifted (expanded) by a lifting factor $L$. 
This works by replacing all zero-entries in the BG by an all-zeros matrix $\matzero_{L\times L}$ and all one-entries by a circularly shifted identity matrix $\bI_L$. The 5G NR standard specifies a set of shift values for each one-entry in the BG, which are selected by the lifting factor $L$. 
The lifted code is thus $L$ times longer than the BG's base-code, i.e., $n_{\textnormal{L}} = Ln_{\textnormal{B}}$ and $k_{\textnormal{L}} = Lk_{\textnormal{B}}$. Since the lifting factor is selected to ensure that $k_{\textnormal{L}} \geq k$, the encoder needs to zero-pad (or fill) the information bit vector $\bmb$ before encoding. {Therefore, $k_{\textnormal{L}} - k$ zeros are appended to $\vecb$}, i.e., $\vecb_{\textnormal{L}}=[\vecb;\veczero_{k_{\textnormal{L}} - k}]$. The appended zeros are called \emph{filler bits}.
To encode $\bmb_{\textnormal{L}}$, the transmitter computes a generator matrix $\tilde\matG$ of size $n_{\textnormal{L}} \times k_{\textnormal{L}}$ from the lifted BG $\tilde\matH$ such that $\tilde\matH\tilde\matG = \matzero_{(n_{\textnormal{L}}-k_{\textnormal{L}})\times k_{\textnormal{L}}}$. The encoder then computes the codeword by $\vecc_{\textnormal{L}}=\textnormal{enc}(\vecb_{\textnormal{L}}) = \tilde\matG \vecb_{\textnormal{L}}$, which {generates} $n_{\textnormal{L}} - k_{\textnormal{L}}$ parity bits {in addition} to the $k_{\textnormal{L}}$ systematic bits in $\vecb_{\textnormal{L}}$.

\begin{figure}[tp]
    \centering
    \includegraphics[width=1.0\columnwidth]{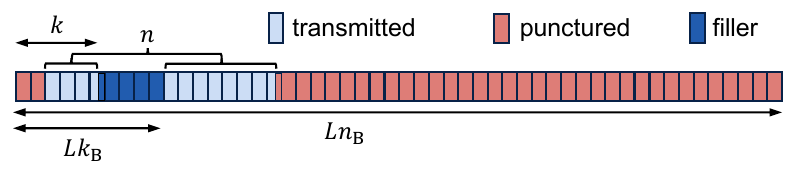}
    \caption{Default 5G NR puncturing pattern for $n=128,~k=64$, working on BG 2 lifted by $L=11$. Each rectangle illustrates a lifted block (i.e., column of the BG). Colors indicate transmitted bits (light blue), punctured bits (red), and filler bits (dark blue) which are always punctured.}
    \label{fig:default_puncturing}
\end{figure}

Third, \emph{puncturing}, which avoids transmission of certain coded bits to match the desired code length $n$.
Since the encoded lifted code-length $n_{\textnormal{L}}$ is in general longer than the desired code-length $n$, the encoder shortens the codeword before transmission. This is achieved by not transmitting (or puncturing) certain bits. 
To decide which bits to puncture, the 5G NR standard specifies the following scheme: (i) the first $2 L$ systematic bits, which correspond to the two highly connected columns in both BGs, are always punctured, (ii) the remaining $k-2L$ systematic bits are transmitted, (iii) the $k_{\textnormal{L}} -k$ filler bits contain no information and are always punctured, (iv) the first $n-k+2L$ parity bits are transmitted, and (v) all remaining parity bits are punctured. 
Thereby, we select $n$ bits out of~$\vecc_{\textnormal{L}}$ to construct the transmit codeword $\vecc$ of desired length~$n$ and rate $r=k/n$.\footnote{The 5G NR standard also defines transmission of additional $n$ parity bits upon unsuccessful decoding, which is called hybrid automatic repeat request (HARQ). As a more powerful LDPC code in the first transmission reduces the frequency of HARQ, we only focus on optimizing the puncturing pattern for the initial (first) transmission. An extension of SPAT for subsequent HARQ transmissions is left for future work.}
An example of this 5G NR LDPC code rate-matching scheme is depicted in \fref{fig:default_puncturing} for $n=128$ and $k=64$, which results in the selection of BG 2 and a lifting factor of $L=11$. 
In contrast to 5G NR compliant puncturing, the proposed SPAT algorithm optimizes the selection of $n$ bits out of the $n_{\textnormal{L}}$ lifted code bits obtained from the encoder.


\subsection{Channel Model}
We transmit each code bit $c_i=[\vecc]_i$ with $i\in\{1,\dots,n\}$ independently over a memoryless BI-AWGN channel. We first map each code bit $c_i$ for transmission to a binary-phase-shift-keying (BPSK) constellation symbol according to $x_i=(-1)^{c_i}$.

Then, we transmit $x_i$ over a real-valued BI-AWGN channel:
\begin{equation}
    y_i = x_i + z_i.
\end{equation}
{Here,} $y_i$ is the channel's output, and $z_i$ is independent and identically distributed (i.i.d.) real-valued Gaussian noise with zero mean and variance $\sigma^2$. We define the SNR as $\textit{SNR}=1/\sigma^2$.

The receiver then uses the channel's outputs $y_i$ to compute a-posteriori probabilities for the transmitted symbol~$x_i$ given~$y_i$, described by the log-likelihood ratio (LLR) value
\begin{equation}
    \ell_i = \log\!\left(\frac{\Prob(x_i = +1|y_i)}{\Prob(x_i = -1|y_i)}\!\right) = \frac{2y_i}{\sigma^2}. \label{eq:llr}
\end{equation}
As the LLR values are computed from the channel output~$y_i$, $\ell_i$ is often called \emph{channel LLR}.
The channel LLRs of the individual code bits $\bm{\ell}=[\ell_1;\dots;\ell_n]$ are then passed to the LDPC MP decoder, which we describe in the following. 

\subsection{MP-Based LDPC Decoding}

Practical LDPC decoders typically utilize MP algorithms that operate on the bipartite factor graph described by the parity-check matrix~\cite{kschischang2001factor}. 
Each row of the parity-check matrix corresponds to a \emph{check node} and each column to a \emph{variable node}.
MP decoders iteratively exchange reliability information (in the form of LLR values) between check and variable nodes according to a predefined schedule.
While the flooding schedule alternates between updating \emph{all} variable nodes and then \emph{all} check nodes in parallel, the layered schedule put forward in~\cite{sharon2007efficient} serially updates blocks (layers) of parity-check equations (i.e., blocks of orthogonal rows of the parity-check matrix). For both schedules, we denote one iteration to involve one update of all parity-check equations and quantify the number of MP iterations with the flooding and the layered schedule by $N_{\textnormal{F}}$ and $N_{\textnormal{L}}$, respectively.
The layered schedule typically requires about half the number of MP iterations to achieve the same error-rate performance as the flooding schedule \cite{sharon2007efficient}. 

Message updates in the variable nodes process the corresponding channel LLR defined in \fref{eq:llr} together with all messages stemming from check nodes connected to this variable node. 
Variable nodes associated with punctured bits assume channel LLR values of zero (absolute uncertainty about the value of a punctured bit); variable nodes associated with filler bits assume channel LLR values of infinity (absolute certainty that filler bits have value zero). 
Message updates in the check nodes can either utilize the sum-product algorithm (SPA) or the max-product algorithm (MPA); other approximations also exist \cite{ren2024generalized}. Each check node processes all messages stemming from the variable nodes connected to this check node in order to produce new messages that are processed by the variable-node updates. 

After a predefined\footnote{To reduce power consumption, early termination strategies are often used to stop decoding as soon as all parity checks are satisfied.} number of MP iterations is reached, the decoder computes the sum of all incoming messages at variable nodes that are associated to systematic bits. The decoder then outputs either the resulting soft-information (representing LLRs) or performs hard decisions based on the sign of this soft-information, to obtain bit estimates for $\vecb$.

\section{Optimization of Puncturing Patterns}
\label{sec:optimization}

We now propose \emph{swapping of punctured and transmitted blocks (SPAT)}, an algorithm that aims at finding better puncturing patterns for MP-based LPDC decoders that carry out a small number of MP iterations. 
In contrast to state-of-the-art puncturing-pattern optimization techniques \cite{LIU20151582,ha2004rate}, SPAT considers a given decoder algorithm with a fixed and practically low number of decoding iterations. 

While an exhaustive search over all possible puncturing patterns would quickly result in prohibitive complexity, e.g., for a short rate-half code of length $n=128$, there already exist $\binom{526}{128}\approx 2.3 \cdot 10^{125}$ possible puncturing patterns. 
SPAT pursues a greedy procedure and only explores puncturing of entire blocks of the lifted BG in order to determine improved puncturing patterns in a computationally tractable manner. Here and in what follows, one block denotes one column of the BG, i.e., $L$ subsequent bits in $\vecc_{\textnormal{L}}$.
More specifically, SPAT has a runtime that scales with the product of the number of transmitted blocks, times the number of punctured blocks, times the computational effort for evaluating the optimization objective. 
For the example in \fref{fig:default_puncturing}, SPAT only explores 371 different puncturing patterns, which is computationally tractable with modern computers.
The overall complexity of SPAT is the number of swaps times the cost of evaluating the optimization objective.

\subsection{Optimization Objective}
The goal of SPAT is to optimize the SNR performance in the waterfall region.
Therefore, SPAT determines the minimum SNR required to achieve a target block error rate (BLER) of, e.g., $\textit{BLER}=10^{-3}$, for each evaluated puncturing pattern. We call the minimum SNR to achieve this target error-rate as the \emph{SNR operating point} abbreviated by $\textit{SNR}$@$\textit{BLER}=10^{-3}$.
Therefore, SPAT samples the BLER at multiple SNR values\footnote{{We sample the BLER by Monte--Carlo simulation. Therefore, we transmit random codewords over the BI-AWGN channel and perform MP decoding with the desired algorithm, schedule, and number of MP iterations.}} and linear interpolates (in logarithmic scale) the BLER vs. SNR curve to extract the SNR operation point where $\textit{BLER}=10^{-3}$. 

If faster runtime is desired, SPAT could be modified to optimize a simpler objective. Alternatively, one could only sample the BLER at one specific SNR and select the puncturing pattern that achieves the lowest BLER value. Note that the third method presented in \cite{LIU20151582} determines the bit error rate (BER) at one specific SNR, which does not necessarily result in the best BLER performance, as pointed out in \cite{Wiesmayr2023}.

\begin{figure}[tp]
    \centering
    \includegraphics[width=.99\columnwidth]{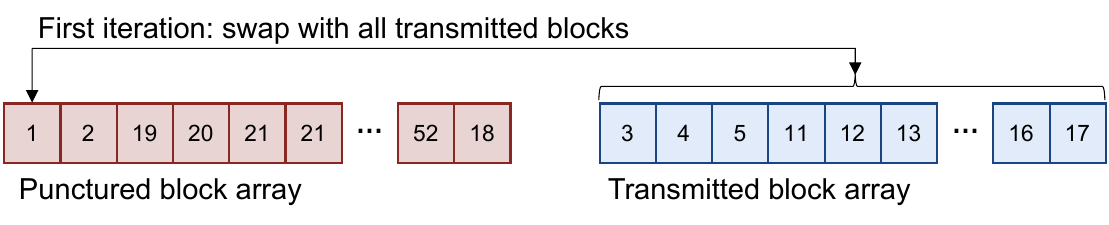}
    \caption{%
    Punctured (left) and transmitted block (right) arrays in the first iteration of \fref{alg:excel} when initialized with the 5G NR puncturing pattern shown in \fref{fig:default_puncturing}. Array elements specify indices of punctured and transmitted blocks, each representing $L$ bits,     except for the last element of the punctured block array, which represents the partially punctured block (cf.\ the 18th block in \fref{fig:default_puncturing}).}
    \label{fig:alg}
\end{figure}

\subsection{SPAT: Swapping of Punctured and Transmitted Blocks}

SPAT proceeds as follows. 
Starting from the 5G NR puncturing pattern, we first store the indices of transmitted and punctured blocks in two arrays \texttt{\small punc\_idx} and \texttt{\small tran\_idx}, respectively. 
For the example in \fref{fig:default_puncturing}, we illustrate the initial array values in \fref{fig:alg}. 
The block that is partially punctured and partially transmitted (block index 18) is defined to be the last one of the punctured block array. 
The block that contains systematic bits and filler bits (block index~6) is for simplicity excluded from the optimization and thereby not listed in both arrays.
The systematic bits of this partial block (block index~6) are always transmitted; the filler bits are always punctured.

SPAT now performs two nested loops. In the outer loop, SPAT iterates over all elements of the punctured block array (\fref{alg:excel} line \ref{eq:alg_outer_loop}).
For each element of the punctured block array, the algorithm benchmarks swapping with all elements of the transmitted block array, one after the other, inside the inner loop (line \ref{eq:alg_inner_loop}).
Within the inner loop, the current punctured block index is swapped with the current transmitted block index (line \ref{eq:exchange}) and the SNR operating point of the resulting temporary puncturing pattern is simulated (line~\ref{eq:sim}). 
After the inner loop has iterated over all transmitted blocks, the block index with the lowest SNR operating point is determined (line \ref{eq:argmin}) and swapped between the two arrays (line~\ref{eq:exchange1}), as this new puncturing patterns showed an improved SNR operating point.
If none of the explored swaps improve the SNR operating point, then the current puncturing pattern is left untouched.
Then, the outer loop proceeds to the next element of the punctured block array and iterates until all are optimized through swapping and testing with all elements of the transmitted block array.

Note that the step on line \ref{eq:alg_sim_first} can be skipped for $\texttt{\small i}>1$, as it corresponds to the previous iteration's best SNR operating point.
Furthermore, note that array indices in \fref{alg:excel} start with value $0$ and puncturing (and transmission) block indices denote columns of the BG and thereby start with value $1$.

Since SPAT optimizes the BLER performance of the actual decoder (line \ref{eq:sim}), we ensure that the resulting puncturing pattern achieves improved few-iteration decoding performance.

\begin{algorithm}[t]
\caption{The SPAT algorithm}
\label{alg:excel}
\begin{algorithmic}[1]
\State {\bfseries input:}
\texttt{\small punc\_idx} and \texttt{\small tran\_idx} 
\For{\texttt{\small i=1} {\bfseries to} \texttt{\small len(punc\_idx)}} \label{eq:alg_outer_loop}
\State \texttt{\small snr\_at\_bler[0] $\gets$ sim(punc\_idx, tran\_idx)} \label{eq:alg_sim_first}
\For{\texttt{\small j=1} {\bfseries to} \texttt{\small len(tran\_idx)}} \label{eq:alg_inner_loop}
\State \texttt{\small punc\_tmp $\gets$ punc\_idx}
\State \texttt{\small tran\_tmp $\gets$ tran\_idx}
\State \texttt{\small swap(punc\_tmp[i-1], tran\_tmp[j-1])} \label{eq:exchange}
\State \texttt{\small snr\_at\_bler[j]\,$\gets$\,sim(punc\_tmp,\,tran\_tmp)} \label{eq:sim}
\EndFor
\State \texttt{\small i\_opt $\gets$ argmin(snr\_at\_bler)} \label{eq:argmin}
\If{\texttt{\small i\_opt > 0}}
\State \texttt{\small swap(punc\_idx[i-1], tran\_idx[i\_opt-1])} \label{eq:exchange1}
\EndIf
\EndFor
\State \textbf{output:~}\texttt{\small punc\_idx} and \texttt{\small tran\_idx}
\end{algorithmic}
\end{algorithm}

\section{Simulation Results} \label{sec:results}

We now demonstrate the efficacy of SPAT for 5G NR LDPC codes. 
To this end, we extend the implementation of the 5G NR LDPC MP decoder from NVIDIA Sionna \cite{Hoydis2022} to support (i) alternative puncturing patterns and (ii) layered MP decoding~\cite{sharon2007efficient}. 
In the following, we present an extensive set of simulation results for various code rates and block lengths.
We start in \fref{sec:bler_vs_snr} by comparing the BLER performance of the 5G-NR-compliant and optimized puncturing patterns and then present in \fref{sec:convergence_bce_analysis} a convergence analysis which compares the BCE for a varying number of MP iterations.

The BLER at each SNR shown in our plots is simulated with $10^6$ codeword transmissions and the SNR gains are compared to a target BLER of $10^{-3}$. The SPAT algorithm simulates each SNR point with $10^6$ codeword transmissions (or until $10^5$ block errors are detected) and stops simulati{ng} higher SNR values when no block error is detected. A block error is detected if at least one of the $k$ information bit estimates at the decoder output is incorrect. 

\subsection{Conventional vs. Optimized Puncturing Patterns}\label{sec:bler_vs_snr}

\begin{figure}[t]
    \centering
    \includegraphics[width=.492\columnwidth]{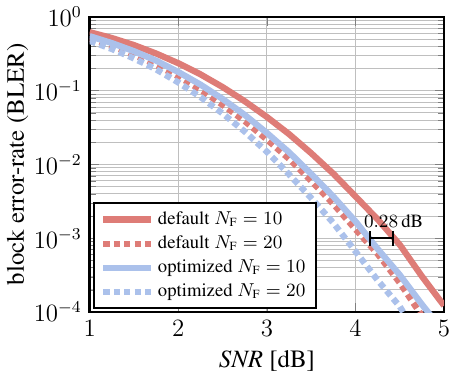}
    \includegraphics[width=.492\columnwidth]{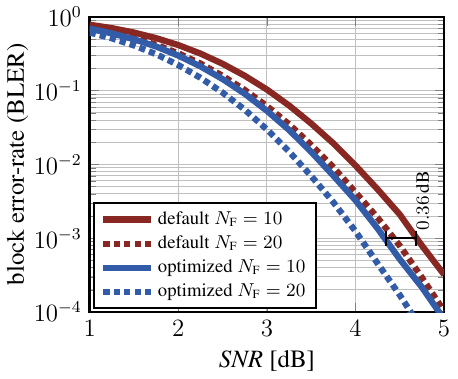}\\
    \includegraphics[width=.492\columnwidth]{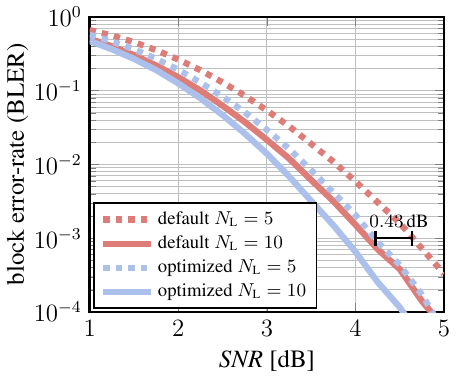}
    \includegraphics[width=.492\columnwidth]{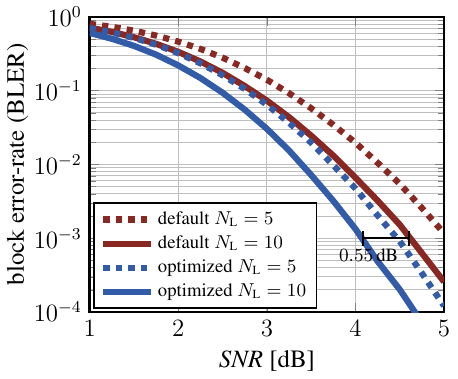}\\
    \includegraphics[width=.98\columnwidth]{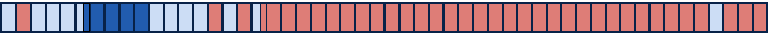}
    \caption{BLER vs. SNR of flooding (top) and layered (bottom) decoding of a rate-half code of length $n=128$ with SPA (left) and MPA (right), puncturing pattern optimized for SPA with $N_{\textnormal{F}}=10$ (bottom).}
    \label{fig:short_code}
\end{figure}

In \fref{fig:short_code}, we consider a short rate-half code of length $n=128$ that is constructed from BG~2. 
Here, we used SPAT for an SPA MP decoder applying the flooding scheme with $N_{\textnormal{F}}=10$ iterations.
SPAT delivers a puncturing pattern that yields a $0.28$\,dB SNR gain compared to the 5G-NR-compliant code (upper left plot). 
This optimized puncturing pattern also achieves $0.36$\,dB lower SNR with a MPA flooding decoder (upper right plot), and $0.43$\,dB and $0.55$\,dB lower SNR with the layered schedule (plots at the bottom of \fref{fig:short_code}), with SPA with $N_{\textnormal{L}}=5$ and MPA with $N_{\textnormal{L}}=10$, respectively. 
Interestingly, the optimized pattern only punctures the second out of the two highly connected blocks, which are considered to be always punctured in the 5G NR standard.

\begin{figure}[t]
    \centering
    \includegraphics[width=.492\columnwidth]{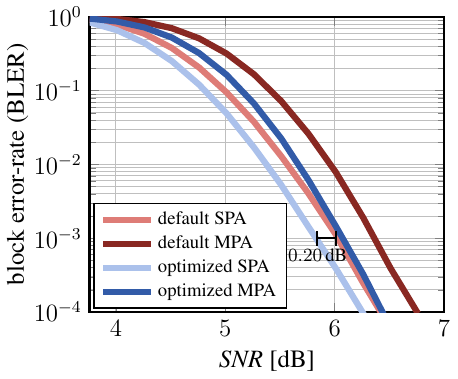}
    \includegraphics[width=.462\columnwidth]{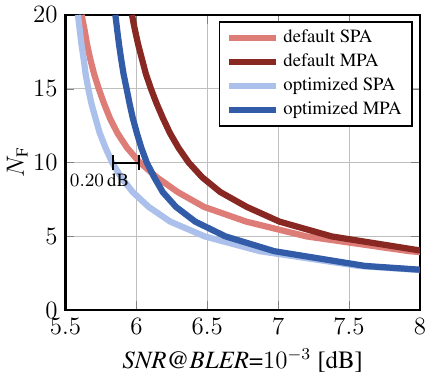}
    \includegraphics[width=.98\columnwidth]{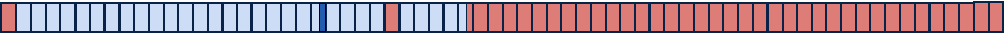}
    \caption{BLER vs. SNR of flooding decoding of a rate-3/4 code of length $n=512$ with $N_{\textnormal{F}}=10$ (left) and SNR operation points for varying $N_{\textnormal{F}}$ (right), puncturing pattern optimized for SPA with $N_{\textnormal{F}}=10$ (bottom).}
    \label{fig:mid_code}
\end{figure}

In \fref{fig:mid_code}, we consider a 
medium-length rate-${3}/{4}$ code of length $n=512$ that is constructed from BG 1. 
Here, we used SPAT for an SPA MP decoder applying the flooding scheme with $N_{\textnormal{F}}=10$ iterations.
SPAT delivers a puncturing pattern that yields a $0.20$\,dB SNR gain compared to the 5G-NR-compliant code (left plot, light colors). 
This optimized puncturing pattern also achieves $0.30$\,dB lower SNR with MPA (left plot, dark colors).
In the right plot of \fref{fig:mid_code}, we compare the SNR operating point for a varying number of flooding iterations $N_{\textnormal{F}}$. Although the puncturing pattern was optimized for $N_{\textnormal{F}}=10$, it shows improved SNR performance compared to the 5G NR standard code up to $N_{\textnormal{F}}=20$ flooding iterations. 
Interestingly, the optimized pattern only punctures the first out of the two highly connected blocks, and punctures the fifth core parity column (fifth block after the filler bits) and many blocks of the extension parity columns (degree-1 variable nodes), as denoted in \cite[Fig.~5]{richardson2018design}.

\begin{figure}[t]
    \centering
    \includegraphics[width=.492\columnwidth]{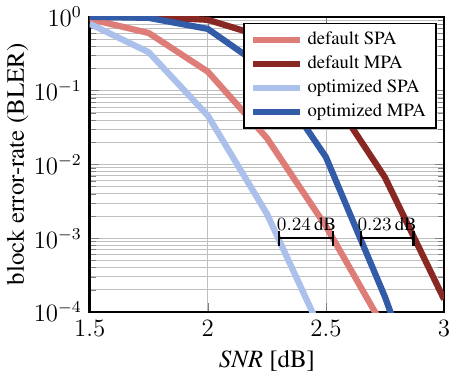}
    \includegraphics[width=.462\columnwidth]{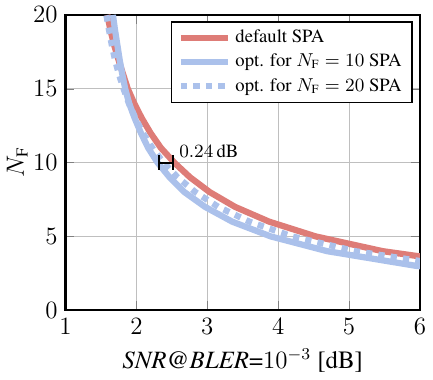}
    \includegraphics[width=.98\columnwidth]{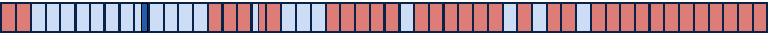}
    \includegraphics[width=.98\columnwidth]{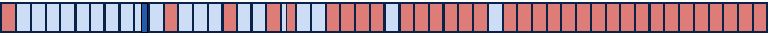}
    \caption{BLER performance of flooding decoding of a rate-half code of length $n=4096$ with $N_{\textnormal{F}}=10$ (left) and SNR operation points for varying $N_{\textnormal{F}}$ (right). Puncturing patterns optimized for SPA with $N_{\textnormal{F}}=10$ (middle) and with $N_{\textnormal{F}}=20$ (bottom).}
    \label{fig:long_code}
\end{figure}

In \fref{fig:long_code}, we consider a long rate-half code of length $n=4096$ that is constructed from BG 2. 
Here, we used SPAT for an SPA MP decoder applying the flooding scheme with $N_{\textnormal{F}}=10$ iterations.
SPAT delivers a puncturing pattern that yields a $0.24$\,dB SNR gain compared to the 5G-NR-compliant code (left plot, light colors). 
This optimized puncturing pattern also achieves $0.23$\,dB lower SNR with MPA (left plot, dark colors).
In the right plot of \fref{fig:long_code}, we compare the SNR operating point for a varying number of flooding iterations $N_{\textnormal{F}}$. 
The puncturing pattern optimized for $N_{\textnormal{F}}=10$ shows improved SNR performance compared to that of the 5G-NR-compliant pattern up to $N_{\textnormal{F}}=16$. Different from our other results shown in \fref{fig:short_code} and \fref{fig:mid_code}, here we also used SPAT to find improved puncturing patterns for 20 instead of 10 flooding iterations. However, the puncturing pattern optimized for $N_{\textnormal{F}}=20$ only shows $0.01$\,dB better SNR performance at $N_{\textnormal{F}}=20$ than the 5G-NR-compliant pattern.  
Interestingly, the pattern optimized for $N_{\textnormal{F}}=10$ punctures both of the two highly connected blocks, but the puncturing pattern optimized for $N_{\textnormal{F}}=20$ only punctures the first one.

This highlights that SPAT can deliver much better puncturing patterns for few-iteration decoding (e.g., for $N_{\textnormal{F}}=10$), while it is challenging to outperform the 5G-NR-compliant puncturing pattern in the long decoding limit (e.g., for $N_{\textnormal{F}}\geq20$).

\subsection{BCE Convergence Analysis}\label{sec:convergence_bce_analysis}

We now take a closer look at the reason why improved puncturing patterns can outperform 5G NR standard-compliant LDPC codes. 
To this end, we compare the empirical average binary cross-entropy (BCE) between the transmitted information bits and the corresponding LLR values at the decoder output for each iteration. 
As shown in \cite{Wiesmayr2023}, the empirical average BCE is a qualitative measure for the BER and it is minimal if the LLR values correspond to ``\emph{true}'' probabilities, according to the definition in \eqref{eq:llr}.
In case of an iterative MP decoder with SPA, we expect that the messages will no longer represent exact likelihoods and can become over- or under-confident after many iterations due to loops in the factor graph.

To compute the empirical average BCE, we first map the $k$ LLR estimates at the decoder output $\hat \ell_i$ to the corresponding probability estimates for $b_i=1$, denoted $\hat p_i$, according to
\begin{equation}
    \hat p_i = \frac{1}{1+e^{\ell_i}}. \label{eq:llr_to_prop}
\end{equation}
We then obtain the average BCE over all information bits by
\begin{equation}
    \textit{BCE} = -\frac{1}{k}\sum_{i=1}^{k} b_i \log(\hat p_i) + (1-b_i) \log(1-\hat p_i),\label{eq:bce}
\end{equation}
which is then averaged over all $10^6$ codeword transmissions at each simulated SNR point, resulting in the empirical average BCE.
To improve numerical stability when calculating \eqref{eq:bce}, we use the equivalent (but numerically stable) form
\begin{equation}
    \textit{BCE} = \frac{1}{k}\sum_{i=1}^{k} \max(0,-\hat\ell_i) + b_i\hat\ell_i + \log\big(1+\exp(-|\hat\ell_i|)\big).\label{eq:bce}
\end{equation}

\begin{figure}[t]
    \centering
    \includegraphics[width=.492\columnwidth]{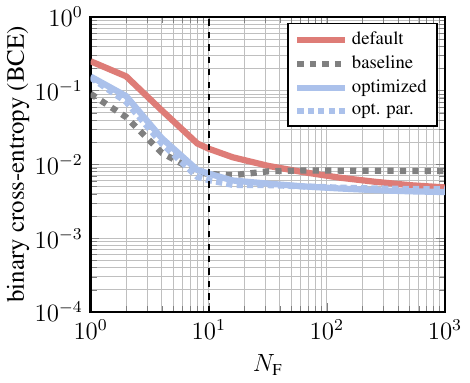}
    \includegraphics[width=.492\columnwidth]{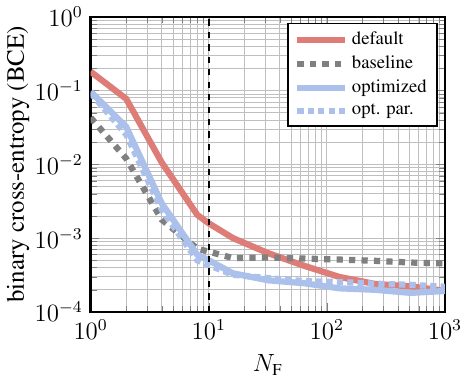}
    \includegraphics[width=.98\columnwidth]{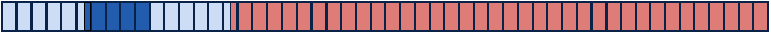}
    \includegraphics[width=.98\columnwidth]{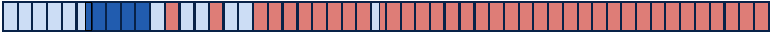}
    \caption{BCE vs. $N_{\textnormal{F}}$ of a rate-half code of length $n=128$, measured at $\textit{SNR}=3$\,dB (left) and $\textit{SNR}=4$\,dB (right). 5G NR compliant puncturing from \fref{fig:default_puncturing} and the optimized puncturing pattern from \fref{fig:short_code} are compared to a baseline (middle) and an optimized parity puncturing pattern (bottom).}
    \label{fig:bce_convergence}
\end{figure}

In \fref{fig:bce_convergence}, we compare the average empirical BCE versus the number of flooding iterations $N_{\textnormal{F}}$ from $1$ to $10^3$. The BCE was measured at an SNR of 3\,dB (left plot) and at an SNR of 4\,dB (right plot). Here, we consider the short rate-half code of length $n=128$ that we also analyzed in \fref{fig:short_code}.
We observed that transmitting (i.e., not puncturing) all information bits leads to a ``decoding head start,'' resulting in lower BCE after the very first iteration compared to patterns which involve puncturing of information bits. 
Due to this observation, we compare the optimized puncturing pattern from \fref{fig:short_code} also with a baseline pattern that we define to transmit the first $n=128$ bits (depicted in \fref{fig:bce_convergence} in the middle) and a pattern optimized with SPAT under the constraint of transmitting all information bits (labeled ``opt. par.,'' depicted in \fref{fig:bce_convergence} at the bottom).
While the baseline puncturing pattern shows the best (lowest) BCE during early iterations, it performs worst after a large number ($64$ or more) of iterations. The 5G-NR-compliant puncturing pattern shows the worst (highest) BCE during early iterations, but shows continuously decreasing BCE up to $N_{\textnormal{F}}=10^3$. Both optimized puncturing patterns perform best from $N_{\textnormal{F}}=8$ and perform approximately as well as the 5G-NR-compliant puncturing pattern at $N_{\textnormal{F}}=10^3$. 

We conclude from this analysis that there is a trade-off between puncturing patterns that work well for the early MP iterations, but then perform not as well for more iterations. 
For this short code of length $n=128$, we observed that the optimized puncturing patterns obtained by SPAT work much better at $N_{\textnormal{F}}=10$ and as well as the 5G-NR-compliant puncturing pattern at $10^3$ MP iterations. However, the results shown in the right plot of \fref{fig:long_code} indicate that optimized puncturing patterns can also perform worse compared to the 5G-NR-compliant pattern, which confirms the fact that the 5G-NR-compliant puncturing pattern was designed for the asymptotic performance limit and that one can do better for few-iteration decoding. 

\begin{figure}[t]
    \centering
    \includegraphics[width=.492\columnwidth]{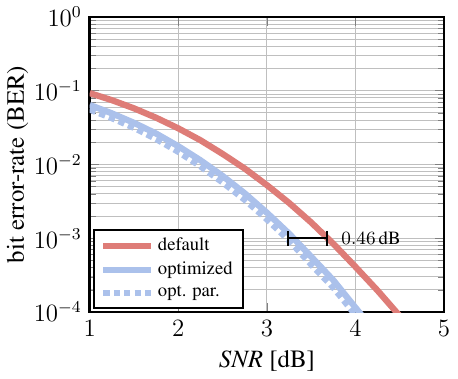}
    \includegraphics[width=.492\columnwidth]{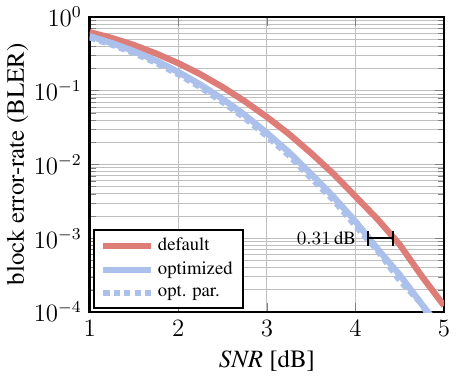}
    \caption{BER (left) and BLER (right) performance of flooding decoding of a rate-half code of length $n=128$ with SPA and $N_{\textnormal{F}}=10$, same puncturing patterns as in \fref{fig:bce_convergence}, which are optimized for SPA with $N_{\textnormal{F}}=10$.}
    \label{fig:ber_bler_short}
\end{figure}

To better understand and quantify the performance difference between the two optimized puncturing patterns compared in \fref{fig:bce_convergence}, we illustrate in \fref{fig:ber_bler_short} their BER (left) and BLER (right) performance with an SPA flooding decoder with $N_{\textnormal{F}}=10$.
Interestingly, the constrained SPAT algorithm obtains a puncturing pattern that performs 0.03\,dB better than the unconstrained SPAT algorithm (cf.~the gain measured in \fref{fig:short_code}).
This minor difference can be explained by the heuristic nature of SPAT, achieving (potentially) different results with different initial puncturing patterns {and optimization constraints}. Note that the constrained SPAT algorithm was initialized with the baseline pattern illustrated in \fref{fig:bce_convergence}.

\section{Conclusions}

We have shown that better puncturing patterns exist for the 5G NR LDPC code when considering practical MP-based decoders {which perform} only a small number of iterations. 
To extract improved puncturing patterns, we have proposed SPAT, a greedy procedure that iteratively swaps blocks of transmitted and punctured bits, and selects puncturing patterns that show better simulated SNR performance. 
By applying SPAT to 5G NR LDPC codes of various lengths and rates, we obtained puncturing patterns that achieve from 0.20\,dB and up to 0.55\,dB better SNR performance than 5G-NR-compliant puncturing in a BI-AWGN channel.
A convergence analysis that is based on the empirical average BCE of bit probability estimates at the decoder output revealed a trade-off between puncturing patterns obtained by SPAT that work best for few-iteration decoding and 5G-NR-compliant puncturing patterns that show in some cases better performance after a large number of iterations.

Since significant SNR gains in few-iteration decoding can be achieved by just transmitting different code bits than those specified by the 5G NR standard, these performance gains come at no additional complexity. 
However, the description of our optimized puncturing patterns is not as simple and universal as that of the 5G-NR-compliant puncturing scheme and also depends on the code length and rate.
Nonetheless, we strongly suggest considering few-iteration decoding performance as code-design criterion in future communication standards. We understand that standardization bodies often formulate code descriptions without consideration of hardware implementation and thereby select a coding scheme that works best in the asymptotic performance limit. Nonetheless, practical decoders operate far from this asymptotic performance limit and{---as we have shown in this paper---}more effective decoding is possible with codes {and puncturing patterns} that are {specifically} designed for the few-iteration decoding regime.

\vspace{.5cm}

\balance


\balance

\end{document}